\documentclass[reprint,amsmath,amssymb,aps,prb,floatfix,twocolumn,superscriptaddress]{revtex4-2}
\usepackage{subfigure}
\usepackage{braket}
\usepackage{lipsum} 
\usepackage[colorlinks=true]{hyperref}
\usepackage{amsmath}
\usepackage{slashed}
\usepackage{xcolor}
\usepackage[normalem]{ulem}
\usepackage{graphicx}
\usepackage{dcolumn}
\usepackage{bm}
\usepackage[version=4]{mhchem}
\usepackage{orcidlink}
\allowdisplaybreaks
\begin{document}

\title{Optical Magnetic Switching in Odd-Parity Magnets with Spin-Orbit Coupling}

\author{SangEun Han\,\orcidlink{0000-0003-3141-1964}
}
\affiliation{Department of Physics and Astronomy, Stony Brook University, Stony Brook, New York 11794-3800, USA}
\author{Qiang Li\,\orcidlink{0000-0002-1230-4832}
}
\affiliation{Department of Physics and Astronomy, Stony Brook University, Stony Brook, New York 11794-3800, USA}
\affiliation{Condensed Matter Physics and Materials Science Division, Brookhaven National Laboratory, Upton, New York 11973-5000, USA}

\date{\today}

\begin{abstract}
    $p$-wave magnets exhibit odd-parity spin polarization in momentum space, with spin splitting that reverses under $\vec{k}\rightarrow -\vec{k}$, while preserving zero net magnetization. 
    Here we show that, in odd-parity magnets with spin-orbit coupling, elliptically polarized light generates a momentum-independent spin-dependent term that dynamically switches a zero-net-magnetization $p$-wave state into a finite spin-polarized state. The Floquet-engineered bands also acquire a nonzero Chern number whose sign is controlled by the light polarization. For $f$-wave magnets, circularly polarized light induces a net out-of-plane magnetization, offering a direct experimental signature.
    Our results establish light as an efficient means of controlling magnetic states, with potential applications in spintronics and quantum information.
\end{abstract}

\maketitle

\emph{Introduction}---
Altermagnetism is a recently identified class of magnetic order with vanishing net magnetization, broken time-reversal symmetry, and momentum-dependent spin splitting in the electronic band structure \cite{Libor2022a,Libor2022b,Bai2024,Song2025}. 
These properties underlie unconventional spin, transport, and topological phenomena, making altermagnets a promising platform for spintronics and emergent quantum effects \cite{Fernandes2024,Fang2024,Ghorashi2024,Rao2024,Antonenko2025,Hadjipaschalis2025}. Altermagnets are usually associated with even-parity spin splitting, such as $d$-, $g$-, and $i$-wave magnetic structures~\cite{Libor2022a}.

Recently, odd-parity counterparts have also been proposed and rapidly developed, including $p$-wave magnets, symmetry classifications of odd-parity magnets, orbital-order-based realizations, microscopic spin models, incommensuration effects, and low-dimensional correlated models \cite{Hellenes2024,Brekke2024,Dsouza2025,Changhee2026,Liu2026,Zhu2026,Luo2025,Zhuang2025,GiBaik2026,Eikeland2026,Li2026,Huang2026}. 
In this case, the spin-resolved bands are split in momentum space, with an odd-parity spin splitting satisfying $\Delta E({\bf k})=-\Delta E(-{\bf k})$, while preserving time-reversal symmetry.
As in conventional altermagnets, odd-parity magnets exhibit zero net magnetization despite spin-split bands in momentum space. However, because the spin splitting is odd under momentum inversion, the corresponding magnetic structures are naturally classified as $p$-, $f$-, or $h$-wave spin-splitting structures \cite{Hellenes2024,Brekke2024,Dsouza2025,Changhee2026}.

Recent work has shown that polarized light can dynamically engineer spin textures and topology in altermagnets with spin-orbit coupling (SOC) \cite{Ghorashi2025}. In altermagnets, elliptically polarized light (EPL) induces a momentum-independent Zeeman-like term as well as additional momentum-dependent SOC terms that are odd in momentum. More generally, the order of the induced SOC is tied to the symmetry of the altermagnetic spin splitting, such that a $k^{2n}$-type splitting gives rise to terms up to order $k^{2n-1}$. Thus, $d$-, $g$-, and $i$-wave altermagnets can generate SOC structures up to linear, cubic, and quintic order, respectively. These light-induced terms can stabilize (nearly) persistent spin textures and modify the low-energy band topology \cite{Schliemann2003,Bernevig2006,Koralek2009,Schliemann2017,Ji2022}.

In this work, we show that odd-parity magnets with SOC exhibit a \emph{distinctly different} Floquet response. Since the magnetic spin splitting is odd in momentum, polarized light generates a leading momentum-independent spin-dependent term. 
This term shifts the spin texture in momentum space and converts the zero-net-magnetization $p$-wave state into a finite spin-polarized state. At the same time, the Floquet-engineered bands acquire a nonzero Chern number whose sign is reversed by changing the light polarization. In the $f$-wave case, circularly polarized light (CPL) eliminates the uniform in-plane component and leaves a purely out-of-plane Fermi-sea-averaged spin polarization.
Our results establish odd-parity magnets as a platform for Floquet control of magnetic and topological properties, with potential applications in optically controlled spintronic and topological devices.

\begin{figure*}
    \centering
    \subfigure[]{\includegraphics[height=0.2\linewidth]{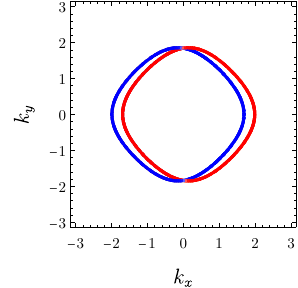}\label{fig:FS1}}
    \subfigure[]{\includegraphics[height=0.2\linewidth]{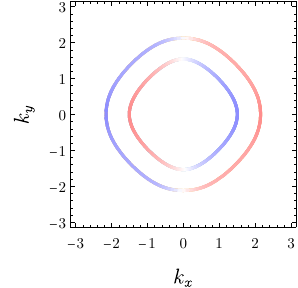}\label{fig:FS2}}
    \subfigure[]{\includegraphics[height=0.2\linewidth]{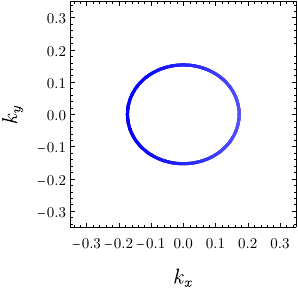}\label{fig:FS3}}
    \subfigure[]{\includegraphics[height=0.2\linewidth]{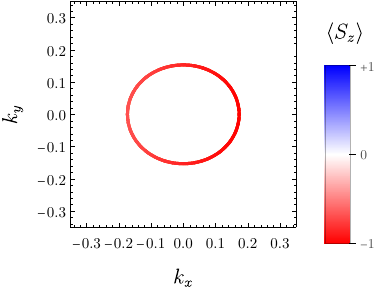}\label{fig:FS4}}
    \caption{The Fermi surfaces before and after introducing polarized light in the square-lattice regularization in the first Brillouin zone. (a) Fermi surface without SOC and light ($A_{x}=A_{y}=0$, $\lambda=0$). (b) Fermi surface with SOC but without light ($A_{x}=A_{y}=0$). (c) Fermi surface with polarized light ($A_{x}=A_{y}=1.8$, $\eta=+1$, and $\lambda=0.6$) and (d) Fermi surface with oppositely polarized light ($A_{x}=A_{y}=1.8$, $\eta=-1$, and $\lambda=0.6$). 
    In (c,d), only the lower band is partially occupied for the chosen chemical potential, and the shown Fermi surfaces are those of the lower band.
    Here, we set $t=1$, $J=0.3$, $\omega=5$, $\theta=0$, and $\mu=2.536$.}
    \label{fig:fermisurface_p_square}
\end{figure*}

\emph{Floquet high-frequency expansion}---
We introduce the polarized light, $\mathbf{A}(t)=(\eta A_{x}\cos(\omega t),A_{y}\sin(\omega t))$. 
For the time-periodic perturbation, we can use the Floquet theorem. Using the Floquet high-frequency expansion, the effective Hamiltonian up to order $1/\omega$ is given by \cite{Oka2009,Kitagawa2011,Eckardt2015,Bukov04032015,Mikami2016,Oka2019,Ghorashi2025,Liu2026,Zhu2026,Lindner2011}
\begin{align}
    H_{\text{eff}}(k)={}&H_{0}+\sum_{n\geq1}\frac{[H_{+n},H_{-n}]}{n\omega}+\mathcal{O}(\omega^{-2}),
    \intertext{where}
    H_{n}={}&\frac{1}{T}\int_{0}^{T}H(t)e^{-in\omega t}dt,
\end{align}
and $T$ is a period of EPL.
For the Peierls-substituted lattice Hamiltonian, the $n=\pm2$ Fourier components have the same form, $H_{+2}=H_{-2}$, and therefore their commutator vanishes, $[H_{+2},H_{-2}]=0$.
Consequently, they do not contribute to the $\mathcal{O}(\omega^{-1})$ effective Hamiltonian, and the leading nonzero spin-dependent Floquet correction is determined by the $n=\pm1$ sector.

\emph{Effective static Hamiltonian of $p$-wave magnet with spin-orbit coupling}---
We start with an effective Hamiltonian for $p$-wave magnet with spin-orbit coupling (SOC) given by \cite{Hellenes2024,Brekke2024},
\begin{align}
    H_{p}={}&t(k_{x}^{2}+k_{y}^{2})\sigma^{0}+\lambda(k_{x}\sigma^{y}-k_{y}\sigma^{x})\notag\\&+J(k_{x}\cos\theta+k_{y}\sin\theta)\sigma^{z},\label{eq:p_hamiltonian}
\end{align}
where $\sigma^{0}$ is the $2\times2$ identity matrix, and $\sigma^{i}$ is the Pauli matrix acting on spin degree of freedom. $\lambda$ and $J$ are coupling constants for the SOC and $p$-wave magnetic terms.
This Hamiltonian is time-reversal symmetric but exhibits momentum-dependent spin-split bands; the momentum-linear $J$ term realizes a minimal $p$-wave spin-splitting structure, yielding zero net magnetization.
Here, we assume the square and triangular lattice regularizations (see End Matter for details). %, $k_{i}\rightarrow\sin(k_{i})$ and $k_{i}^{2}\rightarrow 2(1-\cos(k_{i}))$. 

After applying the Peierls substitution, $k_{x}\rightarrow k_{x}+\eta A_{x}\cos(\omega t)$ and $k_{y}\rightarrow k_{y}+A_{y}\sin(\omega t)$, to the lattice Hamiltonian, where $\eta=\pm1$ is the left/right-handed polarization, we compute $H_{p,\rm eff}$ to leading order in $\omega^{-1}$. The leading low-energy Hamiltonian near the $\Gamma$ point is
\begin{align}
    H_{p,\text{eff}}={}&H_{p}+\lambda'(\cos\theta\sigma^{y}-\sin\theta\sigma^{x})-J'\sigma^{z},\label{eq:eff_p_gamma}
\end{align}
where
\begin{align}
    \lambda'={}&\frac{\eta A_{x}A_{y}J\lambda}{\omega},\quad J'={}\frac{\eta A_{x}A_{y}\lambda^{2}}{\omega}.
\end{align}
A key distinction from light-irradiated even-parity altermagnets with SOC \cite{Ghorashi2025} is that the leading Floquet correction in the $p$-wave case is momentum-independent. 
Polarized light generates an effective uniform spin-dependent field rather than a leading
momentum-dependent SOC term. In the following, we set $\theta=0$ for simplicity; a finite
$\theta$ only rotates the spin-splitting axis and does not change the qualitative conclusions.
This distinction leads to many of the qualitative differences discussed below.

\emph{Fermi surface}---
We present the Fermi surfaces without/with SOC, and with polarized light for the square-lattice regularization in Fig.~\ref{fig:fermisurface_p_square}. The corresponding triangular-lattice results are shown in the End Matter.

The $p$-wave magnet without SOC exhibits clearly spin-split Fermi surfaces, as shown in Fig.~\ref{fig:FS1}.
Each band shows either pure spin-up or spin-down. Due to this, the total magnetization becomes zero. If we introduce the SOC, there is mixing between the spin-up and spin-down bands; however, the total magnetization still remains zero (Fig.~\ref{fig:FS2}).
Polarized light further modifies the spin-split bands through the Floquet-induced uniform spin-dependent field.
For the square-lattice regularization, the Fermi surfaces have the same shape regardless of the polarization, as shown in Figs.~\ref{fig:FS3} and \ref{fig:FS4}. In the case of the triangular-lattice regularization, their Fermi surfaces are distinct for the polarization, as summarized in the End Matter.

\begin{figure}
    \centering
    \subfigure[]{\includegraphics[height=0.25\textwidth]{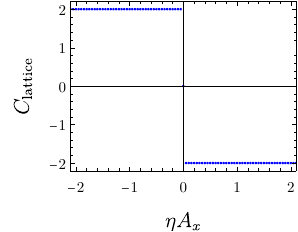}}
    \caption{The lattice Chern number of the lower Floquet-engineered band on a square lattice as a function of $\eta A_{x}$ for fixed $A_{y}$. We set $t=1$, $J=0.3$, $\lambda=0.6$, $\theta=0$, $A_{y}=1.8$, and change $\eta A_{x}$ from $-2$ to $+2$. Thus, CPL occurs at $\eta A_{x}=\pm1.8$. The Chern number of the Floquet-engineered bands changes sign when $\eta A_{x}$ changes sign.}
    \label{fig:chern}
\end{figure}

\emph{Polarization-controlled Chern number}---
Before irradiation, the two bands are not isolated from each other, so a band Chern number is not well defined.
Polarized light opens gaps and generates Floquet-engineered bands with nonzero Chern numbers.
In the $p$-wave case, $C_{\rm lattice}$ denotes the Chern number of the occupied lower Floquet-engineered band.
The upper band carries the opposite Chern number.
With this choice, the Berry curvature near the origin obtained from Eq.~\eqref{eq:eff_p_gamma} is given by
\begin{align}
    \Omega={}&-\frac{\delta\lambda(J^{2}+\lambda^{2})}{2(J^{2}(k_{x}^{2}+\delta^{2})+\lambda^{2}(k_{x}^{2}+k_{y}^{2}+\delta^{2}))^{3/2}},
\end{align}
where $\delta=(\eta A_{x}A_{y}\lambda)/\omega$. 
The local Chern number is given by
\begin{align}
    C_{\text{local}}={}&\frac{1}{2\pi}\iint \Omega dk_{x}dk_{y}=-\frac{\text{sgn}(\lambda\delta)}{2}.
\end{align}
Thus, the sign of the Chern number is controlled by the light polarization
through $\eta A_x A_y$.

By continuing to derive the low-energy effective Hamiltonian near the high-symmetric gap-closing points, we find that for the square lattice, the relevant points are $\Gamma$, $X$, $Y$, and $M$ ($(k_{x},k_{y})=(0,0),(\pi,0),(0,\pi)$, and $(\pi,\pi)$). Each local Berry curvature yields $C_{i}=-\text{sgn}(\eta A_{x}A_{y})/2$ ($i=\Gamma,X,Y,M$), and the sum of the corresponding local Chern numbers reproduces the lattice Chern number, $C_{\text{lattice}}=C_{\Gamma}+C_{X}+C_{Y}+C_{M}=-2\text{sgn}(\eta A_{x}A_{y})$. 
Indeed, if we compute the Chern number over the Brillouin zone, $(k_{x},k_{y})\in (-\pi,\pi)^{2}$, we obtain the same result, $C_{\text{lattice}}=-2\text{sgn}(\eta A_{x}A_{y})$.  The triangular-lattice regularization gives the same result, $C_{\text{lattice}}=-2\text{sgn}(\eta A_{x}A_{y})$, as detailed in the End Matter.
Therefore, in both square and triangular lattice regularizations, changing the sign of $\eta A_{x} A_{y}$ reverses the Chern number of the Floquet-engineered bands. 
The two cases are separated by a gap closing at $\eta A_{x} A_{y}=0$, across which the band Chern number changes from $C_{\text{lattice}}=+2$ to $C_{\text{lattice}}=-2$. 
This demonstrates polarization-controlled topological switching at the level of the Floquet band structure.
Since the system remains metallic, this Chern number would manifest not as a quantized Hall plateau but as a polarization-reversible anomalous Hall response, analogous to the light-induced anomalous Hall effect observed in graphene \cite{Oka2009,McIver2020}. 
This polarization-controlled sign reversal is shown in Fig.~\ref{fig:chern} for the square lattice and in the End Matter for the triangular lattice.

\begin{figure}
    \centering
    \subfigure[]{\includegraphics[width=0.23\textwidth]{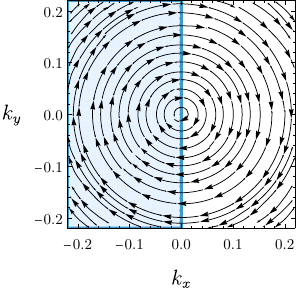}}\hfill
    \subfigure[]{\includegraphics[width=0.23\textwidth]{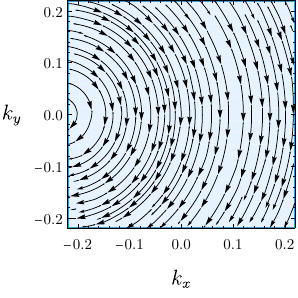}}
    \caption{The spin texture of the lower band of $p$-wave magnet with SOC and (a) without and (b) with polarized light in square lattice regularization. The arrows denote the in-plane spin direction, and the shading denotes the normalized out-of-plane spin polarization. We set $t=1$, $J=0.3$, $\lambda=0.6$, $\theta=0$, and $A_{x,y}=1.8$.}
    \label{fig:spin_texture_p}
\end{figure}
\emph{Spin texture and magnetic structure}---
Let us examine the spin texture of the effective static Hamiltonian. From Eq.~\eqref{eq:eff_p_gamma}, we obtain the light-induced spin-dependent term, $\lambda'(\cos\theta\sigma^{y}-\sin\theta\sigma^{x})-J'\sigma^{z}$. 
When the Floquet-induced spin-dependent terms dominate over the momentum-dependent bare SOC within a small Fermi sea, the spin texture becomes nearly uniform over the occupied states. We present the spin texture for the square-lattice regularization in Fig.~\ref{fig:spin_texture_p}.

\begin{figure}
    \centering
    \subfigure[]{\includegraphics[height=0.18\textwidth]{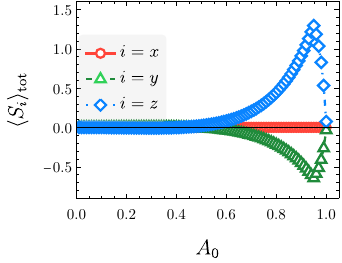}\label{fig:magnetization_p_tot}}\hfill
    \subfigure[]{\includegraphics[height=0.18\textwidth]{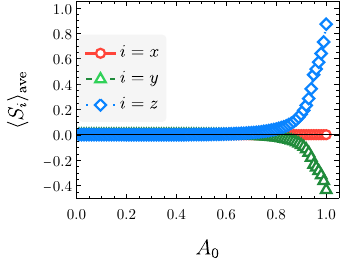}\label{fig:magnetization_p_ave}}
    \caption{The total and averaged spin polarizations over occupied states for the effective static Hamiltonian in the square-lattice regularization. (a) The total spin polarization $\braket{S_{i}}_{\text{tot}}$. (b) Averaged spin polarization $\braket{S_{i}}_{\text{ave}}$. Here we set $t=1$, $J=0.3$, $\lambda=0.6$, $\theta=0$, $\omega=5$, and $A_{x}=A_{y}=1.8A_{0}$, and increase $A_{0}$ from 0 to 1 by 0.01.}
    \label{fig:magnetization_p}
\end{figure}

Also, we compute the total spin polarization over the occupied states, $\braket{S_{i}}_{\rm tot}\equiv\sum_{n}\int_{\rm occ}\frac{d^2k}{(2\pi)^2}
\langle u_{n{\bf k}}|\sigma_i|u_{n{\bf k}}\rangle$, and the averaged spin polarization, $\braket{S_{i}}_{\text{ave}}\equiv\braket{S_{i}}_{\text{tot}}/N_{\text{occ}}$, for the square-lattice regularization, as shown in Fig.~\ref{fig:magnetization_p}. The physical magnetization is proportional to $\braket{S_{i}}_{\rm tot}$. Here, $N_{\rm occ}$ denotes the number of occupied states.
For a fixed chemical potential, increasing the light intensity enhances the spin polarization through the increases of $\lambda'$ and $J'$. However, the total spin polarization does not necessarily increase monotonically. In a certain parameter range, it decreases because the Fermi sea shrinks, reducing the number of occupied states contributing to $\braket{S_{i}}_{\rm tot}$. To separate this effect from the intrinsic spin polarization, we also consider the averaged spin polarization $\braket{S_{i}}_{\text{ave}}$.
For circularly polarized light with a sufficiently small Fermi sea,  the occupied-state average is dominated by the nearly uniform spin texture of the lower band and approaches,
\begin{align}
    \braket{{S}_{x}}_{\text{ave}}={}&\text{sgn}(\eta A_{x}A_{y})\sin\theta\sin\varphi,\\
    \braket{{S}_{y}}_{\text{ave}}={}&-\text{sgn}(\eta A_{x}A_{y})\cos\theta\sin\varphi,\\
    \braket{{S}_{z}}_{\text{ave}}={}&\text{sgn}(\eta A_{x}A_{y})\cos\varphi,
\end{align}
where $\varphi=\tan^{-1}(\lambda/J)$. 
Importantly, the overall sign of the magnetization is determined by the polarization of the polarized light through $\mathrm{sgn}(\eta A_{x}A_{y})$. 
Therefore, if the Fermi surface is sufficiently small, increasing the light intensity drives a dynamical switching from the odd-parity magnetic state to a light-induced spin-polarized state with finite net magnetization.
We present the corresponding results in Fig.~\ref{fig:magnetization_p_ave}.

The origin of the induced magnetization can be understood directly from the structure of the effective Hamiltonian. The light-induced momentum-independent spin-dependent terms shift the center of the spin texture away from the original high-symmetry point in momentum space. Consequently, the contributions from occupied states no longer cancel exactly after integration over the Fermi surface, leading to a finite net magnetization.

This is absent in the even-parity altermagnet. In that case, the light-induced SOC remains odd in momentum and generates a helical or (nearly) persistent spin texture. Since the spin texture remains centered at the high-symmetry point, the momentum-space average of the spin vanishes, and the net magnetization remains zero.

This also provides an estimate for the Fermi-surface size required for a finite magnetization. The induced momentum-independent spin-dependent term displaces the center of the spin texture by $\Delta k\sim\lambda'/\lambda$. A finite magnetization appears only when the Fermi surface remains mostly inside this displaced spin texture region, yielding $k_{F}\lesssim \lambda'/\lambda\approx |A_{x}A_{y}J/\omega|$. This condition provides a simple geometric interpretation of the emergence of the light-induced magnetic response.

\emph{Summary and Discussion}--- In this work, we showed that polarized light can convert a $p$-wave magnet with zero net magnetization into a finite spin-polarized state through a momentum-independent spin-dependent Floquet term. 
The same term renders the Floquet-engineered band structure topologically nontrivial, with a polarization-controlled Chern number.

The induced magnetization originates from the shift of the spin-texture center caused by the uniform Floquet term. When the Fermi surface is sufficiently small, $k_{F}\lesssim |\lambda'/\lambda|\sim |\eta A_{x}A_{y}J/\omega|$, the occupied-state spin polarization no longer cancels over the Fermi sea, producing a finite net magnetization.

An important distinction from previously studied Floquet altermagnets \cite{Ghorashi2025} is that the light-induced correction appears as a momentum-independent spin-dependent term. 
In even-parity altermagnets with $k^{2n}$ spin splitting, polarized light typically generates momentum-dependent SOC terms up to order $k^{2n-1}$, leading to (nearly) persistent spin textures. In contrast, odd-parity $k^{2n+1}$ spin splitting can generate even-parity $k^{2n}$ spin-dependent terms; for the $p$-wave case, this includes a momentum-independent term that acts as an effective uniform spin-dependent field, resulting in finite magnetization and dynamical magnetic switching behavior.

The induced magnetization is not associated with a conventional symmetry breaking transition. Instead, it emerges dynamically through introducing polarized light and can be continuously controlled by the intensity and polarization of the light. Furthermore, both the induced magnetization and the Chern number are controlled by the sign of $\eta A_{x}A_{y}$. 
Consequently, both the induced magnetization and the sign of the band Chern number can be reversed by changing the polarization of the light, providing an all-optical route for controlling magnetism and Floquet band topology.

The triangular-lattice results in the End Matter show the same qualitative behavior, indicating that the response is not tied to a specific lattice regularization. Rather, it relies on the coexistence of odd-parity magnetism and SOC.
Therefore, the light-induced magnetic switching discussed here is expected to be a generic phenomenon in a broad class of odd-parity magnetic systems. In particular, similar Floquet-induced magnetic switching phenomena are expected to arise in a broader class of odd-wave magnets, including higher-angular-momentum counterparts such as $f$-wave magnets. In this sense, the $p$-wave model studied here should be regarded as a representative example of odd-parity magnetic systems rather than a special case \cite{Beaurepaire1996,Stanciu2007,Kirilyuk2010}.

We also consider an $f$-wave odd-parity magnet with SOC. In this case, the uniform in-plane component is proportional to $A_{x}^2-A_{y}^2$ and therefore vanishes for CPL, $|A_{x}|=|A_{y}|$. The remaining in-plane spin texture cancels after Fermi-sea integration, leaving a net out-of-plane magnetization, while EPL can generate both in-plane and out-of-plane components.
This purely out-of-plane response provides a convenient experimental signature, for instance, through polar magneto-optical Kerr measurements~\cite{Sumi2018,Cao2024}.
Moreover, the $f$-wave magnet also develops polarization-controlled Chern bands: for the parameters studied, under CPL, the Floquet-engineered bands carry $C_{\text{lattice}}=\pm2$, and increasing the driving strength induces a sequence of topological transitions accompanied by gap closings (see End Matter).

This observation is particularly encouraging because several recently proposed odd-parity magnetic materials exhibit higher-order momentum-space spin textures. One notable example is \ce{EuAuSb}, which has recently been identified as a helical odd-parity extension of altermagnetism \cite{Sears2025}. 
Our results suggest that such systems may provide realistic platforms for exploring Floquet-controlled magnetization switching and polarization-dependent topological phenomena.

The polarization-dependent control of magnetization and topology provides a route toward all-optical spintronic functionalities \cite{Beaurepaire1996,Stanciu2007,Kirilyuk2010}, enabling magnetic and topological responses to be switched without static magnetic fields. Overall, our results establish odd-parity magnets as a promising platform for Floquet engineering of coupled magnetic and topological phenomena \cite{McIver2020,Wang2013,Shan2021,Kobayashi2023}.

Experimentally, \ce{CeNiAsO}, in which the Ni atoms form a square lattice \cite{Hellenes2024,Yu2025}, and \ce{NiI2}, in which the Ni atoms form a triangular lattice \cite{Song2025b}, have been suggested as candidate materials for realizing $p$-wave magnetism. Our results therefore suggest that irradiating these materials with polarized light could provide a possible route for exploring the predicted Floquet-induced topological and magnetic properties. To estimate the corresponding experimental conditions, we benchmark the dimensionless parameters used throughout the main text against the typical energy scales of these candidate materials. For example, taking $t=0.2\,\mathrm{eV}$ as a representative hopping amplitude, the parameters used in the figures correspond to $\lambda=0.12\,\mathrm{eV}$, $J=0.06\,\mathrm{eV}$, and $\hbar\omega=1.0\,\mathrm{eV}$. This corresponds to a photon wavelength of approximately $1240\,\mathrm{nm}$, placing it in the near-infrared laser regime. Furthermore, the chosen driving parameters, $A_{x}=A_{y}=1.8$ and $\hbar\omega=1.0\,\mathrm{eV}$, correspond to a peak intensity of approximately $\sim 2.69\times10^{12}\,\mathrm{W/cm^2}$ ($E_0\simeq4.5\times10^{9}\,\mathrm{V/m}$ for $a=4\,\text{\AA}$). 
We emphasize that this intensity corresponds to the strong-drive parameters chosen in the figures for illustrative clarity rather than a threshold for the effect. Since $\lambda',J'\propto A_{x}A_{y}/\omega$, both the induced magnetization and the Floquet gap turn on continuously with the driving strength, and their signs, which are the central experimental signatures, are fixed by $\text{sgn}(\eta A_{x}A_{y})$ at any intensity. Moreover, the switching condition $k_{F}\lesssim |A_{x}A_{y}J/\omega|$ can be satisfied at substantially weaker driving for systems with a small Fermi surface. The field strength quoted above lies within the range commonly considered in Floquet-engineering experiments \cite{Wang2013,McIver2020,Shan2021,Kobayashi2023}.

\emph{Acknowledgments}---The work at Stony Brook University was supported by the SUNY Research Foundation for Stony Brook University. Q.~L.~acknowledges the support by the U.S.~Department of Energy, Office of Basic Energy Sciences, Contract No.~DE-SC0012704.

\emph{Data availability}---
The data that support the findings of this article are not publicly available. The data are available from the authors upon reasonable request.

\begin{widetext}
\begin{center}
	\bfseries\MakeUppercase{End Matter}
\end{center}
\end{widetext}
\appendix

\begin{figure*}
    \centering
    \subfigure[]{\includegraphics[height=0.2\linewidth]{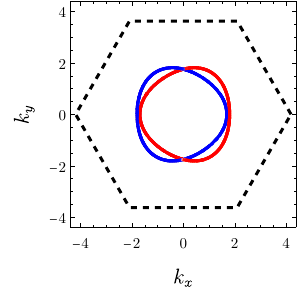}}
    \subfigure[]{\includegraphics[height=0.2\linewidth]{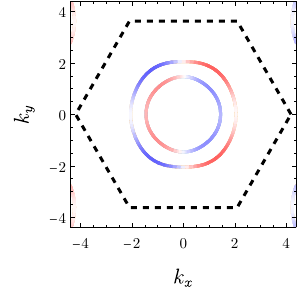}}
    \subfigure[]{\includegraphics[height=0.2\linewidth]{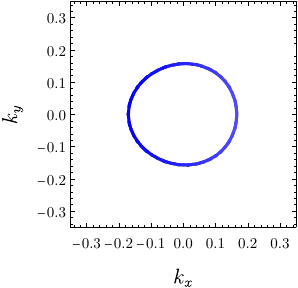}}
    \subfigure[]{\includegraphics[height=0.2\linewidth]{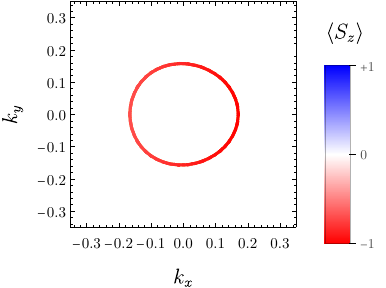}}
    \caption{The Fermi surfaces before and after introducing polarized light in a triangular-lattice regularization in the first Brillouin zone. (a) Without SOC ($\lambda=0$, $A_{x,y}=0$). (b) With SOC ($\lambda=0.6$, $A_{x,y}=0$). (c-d) With polarized light ($\lambda=0.6$, $A_{y}=A_{x}=1.8$, and $\eta=\pm1$ for (c) and (d), respectively). In the light-irradiated cases in (c) and (d), the displayed Fermi surfaces come from the partially occupied lower band, while the upper band remains empty for the chosen chemical potential. We set $t=1,J=0.3,\omega=5$, $\theta=0$, and $\mu=2.536$. The dashed lines in (a) and (b) indicate the first Brillouin zone of the triangular lattice.}
    \label{fig:fermisurface_p_triangular}
\end{figure*}

\begin{figure}
    \centering
    \subfigure[]{\includegraphics[height=0.25\textwidth]{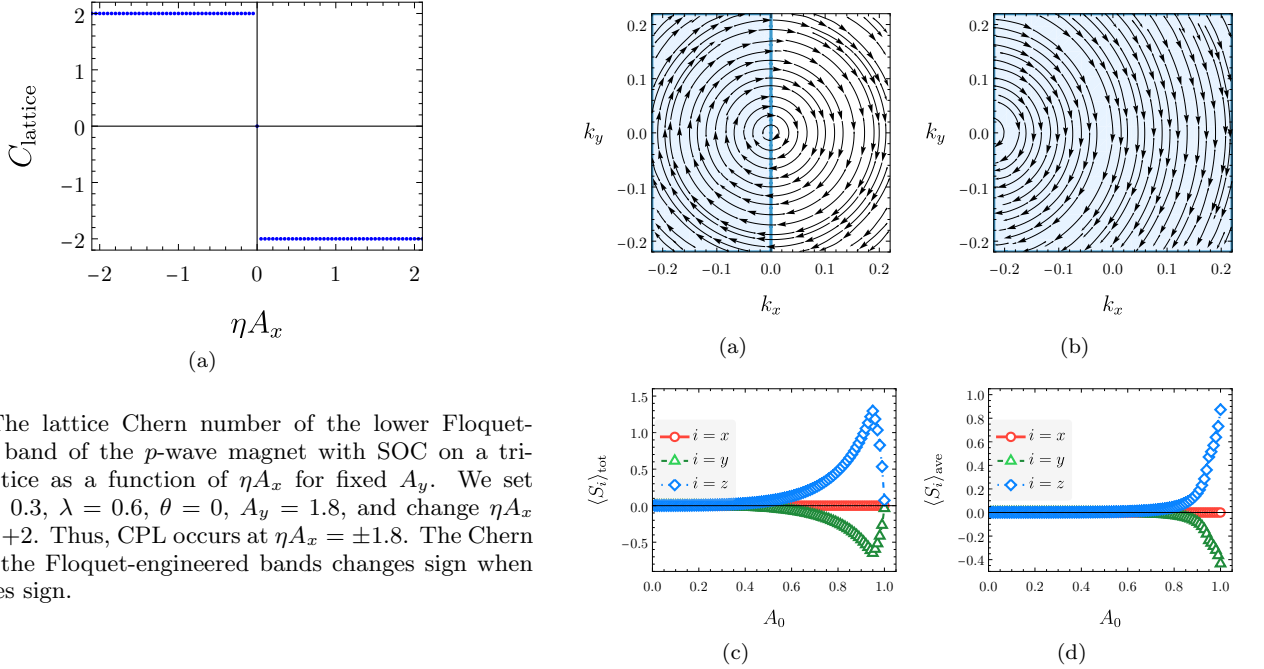}}
    \caption{The lattice Chern number of the lower Floquet-engineered band of the $p$-wave magnet with SOC on a triangular lattice as a function of $\eta A_{x}$ for fixed $A_{y}$. We set $t=1$, $J=0.3$, $\lambda=0.6$, $\theta=0$, $A_{y}=1.8$, and change $\eta A_{x}$ from $-2$ to $+2$. Thus, CPL occurs at $\eta A_{x}=\pm1.8$. The Chern number of the Floquet-engineered bands changes sign when $\eta A_{x}$ changes sign.}
    \label{fig:chern_p_triangular}
\end{figure}

\begin{figure}
    \centering
    \subfigure[]{\includegraphics[width=0.23\textwidth]{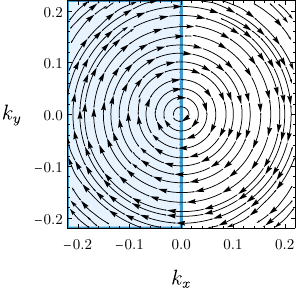}}\hfill
    \subfigure[]{\includegraphics[width=0.23\textwidth]{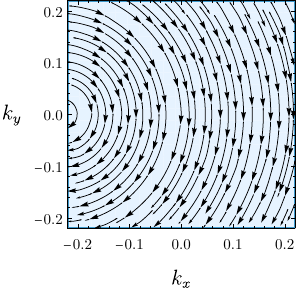}}\\
    \subfigure[]{\includegraphics[height=0.18\textwidth]{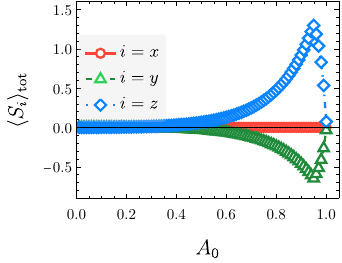}\label{fig:magnetization_p_tot_3}}\hfill
    \subfigure[]{\includegraphics[height=0.18\textwidth]{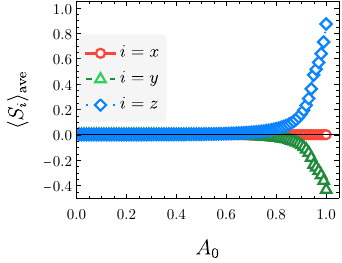}\label{fig:magnetization_p_ave_3}}
    \caption{The triangular-lattice results for the $p$-wave magnet with SOC. (a,b) The spin textures of the lower band (a) without and (b) with polarized light. The arrows denote the in-plane spin direction, and the shading denotes the normalized out-of-plane spin polarization. (c,d) The spin polarizations over occupied states: (c) total spin polarization $\braket{S_i}_{\rm tot}$ and (d) averaged spin polarization $\braket{S_i}_{\rm ave}$. For (a,b), we set $t=1$, $J=0.3$, $\lambda=0.6$, $\theta=0$, and $A_{x}=A_{y}=1.8$; for (c,d), we set $t=1$, $J=0.3$, $\lambda=0.6$, $\theta=0$, and $\omega=5$, and increase $A_{x}=A_{y}$ from 0 to 2 in steps of 0.05.}
    \label{fig:spin_texture_polarization}
\end{figure}

\emph{Lattice regularization}---\label{end:lattice}
Here, we will discuss the lattice regularization schemes used in the main text. We start with the effective Hamiltonian for the $p$-wave magnet with SOC, in Eq.~\eqref{eq:p_hamiltonian}. To perform the Peierls substitution, we need a lattice regularization, which is a lattice model with Eq.~\eqref{eq:p_hamiltonian} as its effective model. It can be obtained in diverse lattice structures, but here, we will consider square and triangular lattices.

\begin{figure*}[ht]
    \centering
    \subfigure[]{\includegraphics[height=0.12\textwidth]{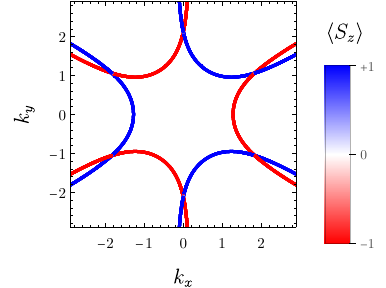}}
    \subfigure[]{\includegraphics[height=0.12\textwidth]{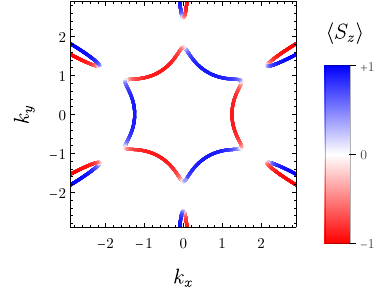}}
    \subfigure[]{\includegraphics[height=0.12\textwidth]{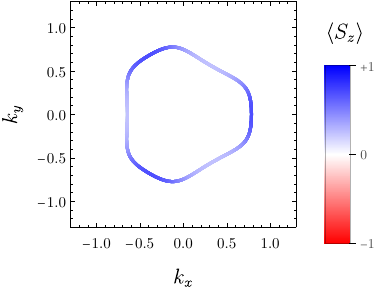}}
    \subfigure[]{\includegraphics[height=0.12\textwidth]{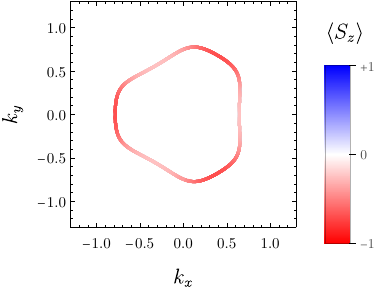}}
    \subfigure[]{\includegraphics[height=0.12\textwidth]{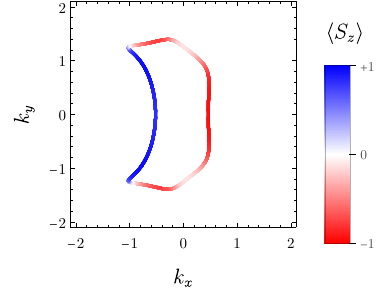}}
    \subfigure[]{\includegraphics[height=0.12\textwidth]{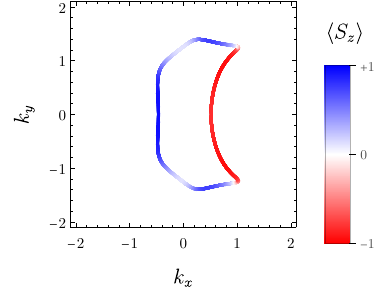}}
    \caption{The Fermi surfaces of $f$-wave magnet before and after introducing polarized light in triangular-lattice regularization in the first Brillouin zone. (a) Without SOC ($\lambda=0$, $A_{x,y}=0$). (b) With SOC ($\lambda=0.8$, $A_{x,y}=0$). (c-d) With CPL ($\lambda=0.8$, $A_{y}=A_{x}=1.8$, and $\eta=\pm1$ for (c) and (d), respectively). (e-f) With EPL ($\lambda=0.8$, $A_{x}=1.8$, $A_{y}=0.9$, and $\eta=\pm1$ for (e) and (f), respectively). For the light-irradiated cases in (c-f), the lower band is fully occupied, and the displayed Fermi surfaces correspond to the partially occupied upper band. We set $t=1,J=1,\omega=3$, $\theta=0$, and $\mu=3.37$. 
    }
    \label{fig:fermisurface_f_triangular}
\end{figure*}

First of all, we consider the square lattice regularization. Let us consider a square lattice with a lattice constant $a$, having $\mathbf{a}_{1}=(a,0)$ and $\mathbf{a}_{2}=(0,a)$. Then, from the nearest hopping
\begin{align}
H={}&4t\sum_{\mathbf{r}}c_{\mathbf{r}}^{\dagger}c_{\mathbf{r}}-t\sum_{\braket{ij}}c_{i}^{\dagger}c_{j}-\frac{i\lambda}{2}\sum_{\braket{ij}}c_{i}^{\dagger}[\bm{\sigma}\cdot(\hat{\mathbf{d}}_{ij}\times\hat{z})]c_{j}
    \notag\\
    &-\frac{iJ}{2}\sum_{\braket{ij}}c_{i}^{\dagger}[\sigma^{z}(\hat{\mathbf{n}}\cdot\hat{\mathbf{d}}_{ij})]c_{j},
\end{align}
where $\hat{\mathbf{d}}_{ij}=(\mathbf{r}_{i}-\mathbf{r}_{j})/a$, $\hat{\mathbf{d}}_{ij}=-\hat{\mathbf{d}}_{ji}$, and $\hat{\mathbf{n}}=(\cos\theta,\sin\theta)$.
In the momentum space, the Hamiltonian can be written
\begin{align}
    H(\mathbf{k})={}&2t[2-\cos(k_{x}a)-\cos(k_{y}a)]\sigma^{0}\notag\\&+\lambda[\sin(k_{x}a)\sigma^{y}-\sin(k_{y}a)\sigma^{x}]\notag\\&+J[\sin(k_{x}a)\cos\theta+\sin(k_{y}a)\sin\theta]\sigma^{z}.\label{eq:squa_lattice_k}
\end{align}
Near the $\Gamma$ point, it has the effective Hamiltonian shown in Eq.~\eqref{eq:p_hamiltonian}.

\begin{figure}
    \centering
    \subfigure[]{\includegraphics[width=0.33\linewidth]{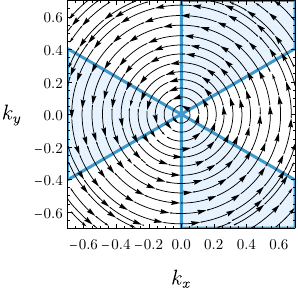}}\hfill
    \subfigure[]{\includegraphics[width=0.33\linewidth]{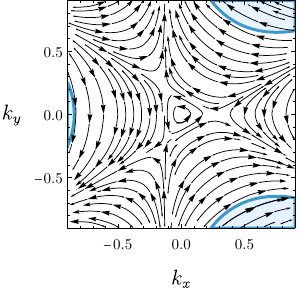}}\hfill
    \subfigure[]{\includegraphics[width=0.33\linewidth]{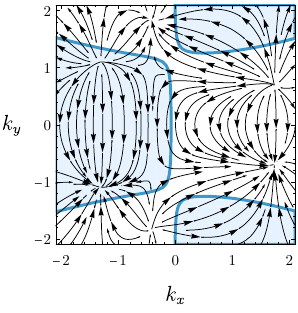}}
    \caption{The spin texture of the upper band near $\Gamma$ point before and after introducing polarized light. (a) without polarized light. (b) with CPL ($A_{x}=A_{y}=1.8$). Near the Gamma point, out-of-plane spin weakly depends on the momentum but in-plane spin depends on the momentum. (c) with EPL ($A_{x}=1.8$ and $A_{y}=0.9$). Near the $\Gamma$ point, both in-plane and out-of-plane spin components show nearly persistent spin texture. We set $t=J=1$, $\omega=3$, and $\lambda=0.8$.}
    \label{fig:spin_texture_f}
\end{figure}

Second, we consider the triangular lattice regularization. Let us consider the triangular lattice with $\mathbf{a}_{1}=(a,0)$, $\mathbf{a}_{2}=(a/2,\sqrt{3}a/2)$, and $\mathbf{a}_{3}=(-a/2,\sqrt{3}a/2)$.
\begin{align}
    H={}&4t\sum_{i}c_{i}^{\dagger}c_{i}-\frac{2t}{3}\sum_{\braket{ij}}c_{i}^{\dagger}c_{j}
    -\frac{i\lambda}{3}\sum_{\braket{ij}}c_{i}^{\dagger}[\bm{\sigma}\cdot(\hat{d}_{ij}\times\hat{z})]c_{j}\notag\\&
    -\frac{iJ}{3}\sum_{\braket{ij}}c_{i}^{\dagger}[\sigma^{z}(\hat{\mathbf{n}}\cdot\hat{\mathbf{d}}_{ij})]c_{j},
\end{align}
in the momentum space,
\begin{align}
    H(\mathbf{k})={}&t\epsilon(\mathbf{k})\sigma^{0}
    +\lambda\left[f_{y}(\mathbf{k})\sigma^{y}-f_{x}(\mathbf{k})\sigma^{x}\right]\notag\\&
    +J\left[f_{x}(\mathbf{k})\cos\theta +f_{y}(\mathbf{k})\sin\theta \right]\sigma^{z},\label{eq:tri_lattice_k}
\end{align}
where
\begin{align}
    \epsilon(\mathbf{k})={}&\frac{3}{4}\left[3-\cos(k_{x}a)-2\cos\bigl(\tfrac{k_{x}a}{2}\bigl)\cos\bigl(\tfrac{\sqrt{3}k_{y}a}{2}\bigl)\right],\\
    f_{x}(\mathbf{k})={}&\frac{2}{3}\left[\sin(k_{x}a)+\sin\bigl(\tfrac{\sqrt{3}k_{y}a}{2}\bigl)\right],\\
    f_{y}(\mathbf{k})={}&\frac{2}{\sqrt{3}}\sin\bigl(\tfrac{\sqrt{3}k_{y}a}{2}\bigl)\cos\bigl(\tfrac{k_{x}a}{2}\bigl).
\end{align}
Near the $\Gamma$ point, it has the effective Hamiltonian shown in Eq.~\eqref{eq:p_hamiltonian}.

In the main text, we used Eq.~\eqref{eq:tri_lattice_k} and Eq.~\eqref{eq:squa_lattice_k} for the Peierls substitution.

\emph{Details of High-frequency expansion}\label{end:fourier}--- 
After performing the Peierls substitution in the square-lattice-regularized Hamiltonian in $p$-wave magnet with SOC and computing the Fourier components, the leading terms are
\begin{align}
    H_{p,0}={}&2t[2-J_{0}(\eta A_{x})\cos(k_{x})-J_{0}(A_{y})\cos(k_{y})]\sigma^{0}\notag\\
    &+\lambda[J_{0}(\eta A_{x})\sin(k_{x})\sigma^{y}-J_{0}(A_{y})\sin(k_{y})\sigma^{x}]\notag\\
    &+J[J_{0}(\eta A_{x})\sin(k_{x})\cos\theta+J_{0}(A_{y})\sin(k_{y})\sin\theta]\sigma^{z}\\
    H_{p,\pm1}={}&2t[J_{1}(\eta A_{x})\sin(k_{x})\mp iJ_{1}(A_{y})\sin(k_{y})]\sigma^{0}\notag\\
    &+\lambda(J_{1}(\eta A_{x})\cos(k_{x})\sigma^{y}\pm iJ_{1}(A_{y})\cos(k_{y})\sigma^{x})\notag\\
    &+J[J_{1}(\eta A_{x})\cos(k_{x})\cos\theta\mp iJ_{1}(A{y})\cos(k_{y})\sin\theta]\sigma^{z}
\end{align}
where we set a lattice constant $a=1$.
The $n=\pm2$ Fourier components satisfy $H_{+2}=H_{-2}$ and therefore they give no $O(\omega^{-1})$ correction, because $[H_{+2},H_{-2}]=0$.
Using $H_{p,0}$ and $H_{p,\pm1}$ above, the effective static Hamiltonian is given by
\begin{align}
H_{p,\text{eff}}={}&2t[2-J_{0}(\eta A_{x})\cos(k_{x})-J_{0}(A_{y})\cos(k_{y})]\sigma^{0}\notag\\
&+\lambda[J_{0}(\eta A_{x})\sin(k_{x})\sigma^{y}-J_{0}(A_{y})\sin(k_{y})\sigma^{x}]\notag\\
&+J[J_{0}(\eta A_{x})\cos\theta\sin(k_{x})+J_{0}(A_{y})\sin\theta\sin(k_{y})]\sigma^{z}\notag\\
&+\frac{4J\lambda J_{1}(\eta A_{x})J_{1}(A_{y})}{\omega}\notag\\
&\qquad \times\cos(k_{x})\cos(k_{y})(\cos\theta\sigma^{y}-\sin\theta\sigma^{x})\notag\\
&-\frac{4\lambda^{2}J_{1}(\eta A_{x})J_{1}(A_{y})}{\omega}\cos(k_{x})\cos(k_{y})\sigma^{z}.
\end{align}
Expanding this around $\Gamma$ point with $J_{0}(x)\approx 1$ and $J_{1}(x)\approx x/2$ for sufficiently small $x$, we obtain the low-energy effective static Hamiltonian in Eq.~\eqref{eq:eff_p_gamma}.

\emph{Triangular-lattice results}--- 
We also performed the same analysis for the triangular-lattice regularization in Eq.~\eqref{eq:tri_lattice_k}. 
The resulting Fermi surfaces are shown in Fig.~\ref{fig:fermisurface_p_triangular}. 
As in the square-lattice case, polarized light modifies the spin-split bands through the Floquet-induced uniform spin-dependent field. 
The lattice Chern number also changes sign under reversal of $\eta A_{x}A_{y}$, confirming the polarization-controlled Chern-band topology in the triangular lattice, as shown in Fig.~\ref{fig:chern_p_triangular}. 
Similar to the square-lattice regularization results in the main text, for the triangular lattice, there are four gap-closing points, $\Gamma$ and $M_{1,2,3}$, and each gap-closing point contributes $C_{\text{local}}=-\text{sgn}(\eta A_{x}A_{y})/2$, leading again to $C_{\text{lattice}}=-2\text{sgn}(\eta A_{x}A_{y})$.
The spin texture and the total and averaged spin polarizations show the same qualitative behavior as in the square-lattice regularization, as shown in Fig.~\ref{fig:spin_texture_polarization}.

\begin{figure}[t]
    \centering
    \subfigure[]{\includegraphics[height=0.18\textwidth]{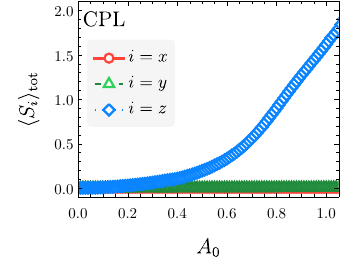}}
    \subfigure[]{\includegraphics[height=0.18\textwidth]{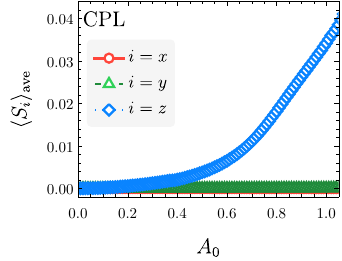}}
    \subfigure[]{\includegraphics[height=0.18\textwidth]{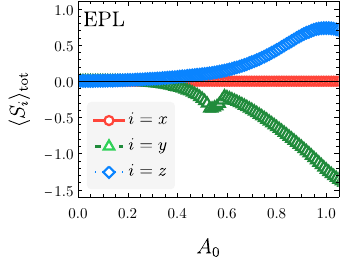}}
    \subfigure[]{\includegraphics[height=0.18\textwidth]{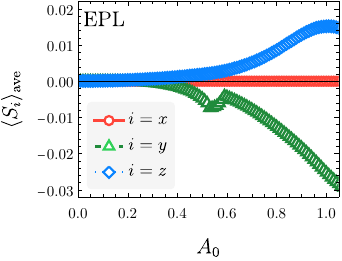}}
    \caption{The spin polarizations in $f$-wave magnet with (a-b) CPL and (c-d) EPL. (a-b) The total and averaged spin polarizations with CPL ($A_{x}=A_{y}=1.8A_{0}$). CPL induces a finite out-of-plane magnetization at large $A_{0}$. (c-d) The total and averaged spin polarization with EPL ($A_{x}=1.8A_{0}$ and $A_{y}=0.9A_{0}$). EPL induces both in-plane and out-of-plane spin polarizations.}
    \label{fig:f_magnetization}
\end{figure}

\emph{Estimation for physical parameters}\label{end:esti}--- 
In the plots of the main text, we use the parameters $t=1$, $J=0.3$, $\lambda=0.6$, $\omega=5$, and $A_{x}=A_{y}=1.8$. We use $a=4\,\text{\AA}$ as a representative in-plane lattice constant for the candidate systems, \ce{NiI2} and \ce{CeNiAsO}.
We assume that $t=1$ corresponds to $0.2\,\rm{eV}$. Then $J=0.06\,\rm{eV}$, $\lambda=0.12\,\rm{eV}$, and $\hbar\omega=1.0\,\rm{eV}$. In particular, this corresponds to a photon wavelength of $\sim1240\,\rm{nm}$, which lies in the near-infrared (NIR) regime.

For the light amplitude $A_{x,y}$, since it is derived from the Peierls substitution, the dimensionless vector potential can be related to the physical vector potential through
\begin{align}
    A_{x,y}=\frac{ea}{\hbar}A_{x,y}^{(0)},
\end{align}
where $A_{x,y}^{(0)}$ is the physical vector potential. Thus,
\begin{align}
    \mathbf{A}^{(0)}(t)=\frac{\hbar}{ea}(\eta A_{x}\cos\omega t,A_{y}\sin\omega t).
\end{align}
The electric field is obtained from $\mathbf{E}(t)=-\partial_t\mathbf{A}^{(0)}(t)$, so the amplitude of the electric field is $E_0=\omega A^{(0)}$. Using the above parameters, we obtain $E_{0}\simeq 4.5\times 10^{9}\,\rm{V/m}$.
The corresponding laser intensity is
\begin{align}
    I=\frac{c\epsilon_{0}}{2}E_{0}^{2}\sim 2.69\times10^{12}\,\rm{W/cm^2},
\end{align}
where we use $c\approx3.0\times10^{8}\,\rm{m/s}$ and $\epsilon_{0}\approx8.85\times10^{-12}\,\rm{F/m}$.

\begin{figure}[ht]
    \centering
    \subfigure[]{\includegraphics[height=0.3\textwidth]{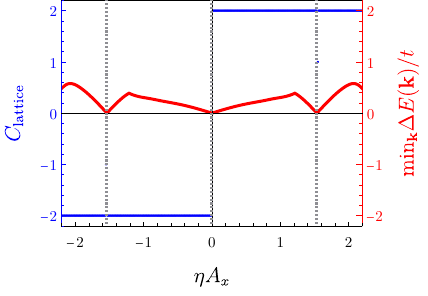}}
    \subfigure[]{\includegraphics[height=0.3\textwidth]{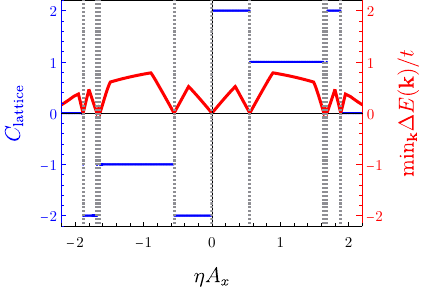}}
    \caption{The lattice Chern number $C_{\text{lattice}}$ (blue, left axis) of the upper band and minimum band gap $\min_{\mathbf{k}}\Delta E(\mathbf{k})/t$ (red, right axis) of the Floquet-engineered $f$-wave magnet with spin-orbit coupling on a triangular lattice as functions of $\eta A_{x}$ for (a) $A_{y}=0.9$ and (b) $A_{y}=1.8$. We set $t=J=1$, $\lambda=0.8$, $\omega=3$, and $\theta=0$.
    Gray dashed lines mark gap-closing points. In (a), only the gap closing at $\eta A_{x}=0$ changes $C_{\rm lattice}$, while the other gap closings leave the Chern-number plateau unchanged.}
    \label{fig:f_chern}
\end{figure}

We note that the high-frequency expansion is controlled by the ratio of the internal energy scales of the low-energy model to the photon energy; for the parameters used in the $p$-wave calculations, $J/\omega=0.06$, $\lambda/\omega=0.12$, and $\mu/\omega\approx0.5$ are all below unity. Since the system remains metallic, heating from the drive is ultimately unavoidable; however, in the high-frequency regime, the energy absorption requires multiphoton many-body processes and is exponentially suppressed, so that the effective static description of Eq. (4) applies within the prethermal window \cite{Kuwahara2016,Mori2016,Abanin2017}. In practice, such conditions are realized transiently in pump–probe setups \cite{Wang2013,McIver2020}, which further mitigates heating.

\emph{$f$-wave magnet with SOC under EPL/CPL}--- 
The effective Hamiltonian of the $f$-wave magnet with SOC is given by
\begin{align}
    H_{f}={}&t(k_{x}^{2}+k_{y}^{2})\sigma^{0}+\lambda(k_{x}\sigma^{y}-k_{y}\sigma^{x})\notag\\&
    +J(k_{y}(3k_{x}^{2}-k_{y}^{2})\cos\theta+k_{x}(3k_{y}^{2}-k_{x}^{2})\sin\theta)\sigma^{z}.
\end{align}
In the $f$-wave case, we use the triangular-lattice regularization.
Here, we obtain the effective static Hamiltonian from the leading $\mathcal{O}(\omega^{-1})$ Floquet high-frequency expansion,
\begin{widetext}
\begin{align}
    H_{f,\text{eff}}={}&H_{f}+\lambda'[-\frac{3}{4}(A_{x}^{2}-A_{y}^{2})(\cos\theta\sigma^{x}+\sin\theta\sigma^{y})-3(k_{+}^{2}\sigma^{+}+k_{-}^{2}\sigma^{-})\cos\theta+3i(k_{+}^{2}\sigma^{+}-k_{-}^{2}\sigma^{-})\sin\theta]-J'\sigma^{z},
\end{align}
\end{widetext}
where $\lambda'$ and $J'$ are defined in the main text, $k_{\pm}\equiv k_{x}\pm ik_{y}$, and $\sigma^{\pm}\equiv(\sigma^{x}\pm i\sigma^{y})/2$.
The Fermi surface before and after polarized-light introducing is presented in Fig.~\ref{fig:fermisurface_f_triangular}.
A momentum-independent spin-dependent term arises in the $f$-wave case as well. Its uniform in-plane component is proportional to $A_{x}^2-A_{y}^2$, and therefore vanishes for CPL, $|A_{x}|=|A_{y}|$, while it remains finite for EPL, $|A_{x}|\neq |A_{y}|$. As a result, CPL leaves a net out-of-plane magnetization after Fermi-sea integration, whereas EPL can generate both in-plane and out-of-plane magnetizations, as shown in Figs.~\ref{fig:spin_texture_f} and~\ref{fig:f_magnetization}.

Furthermore, the $f$-wave magnet develops nontrivial Chern-band topology under
polarized light, as shown in Fig.~\ref{fig:f_chern}. We set $t=J=1$,
$\lambda=0.8$, $\omega=3$, and $\theta=0$, and vary $\eta A_{x}$ for fixed
$A_{y}$. For these parameters, $J/\omega\simeq0.33$ and
$\lambda/\omega\simeq0.27$ remain below unity, so the high-frequency expansion
is still controlled.

For $A_{y}=0.9$, the Floquet-engineered bands carry
$C_{\rm lattice}=\pm2$, with the sign determined by the light polarization.
In this case, $C_{\rm lattice}$ changes only when the sign of $\eta A_{x}$ is
reversed. Additional gap closings occur at finite $\eta A_{x}$, but they do not
change the Chern-number plateau.

For $A_{y}=1.8$, the system undergoes additional topological phase transitions
as $|\eta A_{x}|$ is varied. Specifically, the system has
$|C_{\rm lattice}|=2$ for $|\eta A_{x}|<0.555$, changes to
$|C_{\rm lattice}|=1$ after the gap closing at $|\eta A_{x}|\simeq0.555$,
becomes trivial for $1.639<|\eta A_{x}|<1.683$, returns to
$|C_{\rm lattice}|=2$ for $1.683<|\eta A_{x}|<1.887$, and becomes trivial again
for $|\eta A_{x}|>1.887$. Thus, the Floquet drive provides a direct route to
tuning the Chern-band topology of the $f$-wave magnet.

\end{document}